# Plunging plasma blobs near the marginally stable orbit of Sgr A*.

E M Howard[1]

[1]Department of Physics and Astronomy, Macquarie University, Sydney, 2109, Australia

**Abstract** Multi-wavelength monitoring of Sgr A* flaring activity confirms the presence of embedded structures within the disk on size scales commensurate with the innermost accretion region, matching size scales that are derived from observed light curves within a broad range of wavelengths. We explore here a few of the observational signatures for an orbiting spot in non-keplerian motion near the event horizon of Sgr A* and model light curves from plunging emitting material near the marginally stable orbit of Sgr A*. All special and general relativistic effects (relativistic beaming, redshifts and blue-shifts, lensing effect, photon time delays) for unpolarized synchrotron emission near a Schwarzschild and Kerr black hole are all taken into consideration.

## 1. Introduction

In the past few years, observations of emission from accreting black holes have introduced new possibilities for astrophysical tests of relativistic effects. Observations of AGNs at various frequencies indicate the presence of time variability in the flare emission. The origin of this variation is not yet understood. Recent discoveries provide direct evidence for strong gravitational signatures in supermassive black holes with the prospect of determining the black hole's spin and the nature of the variable emission. The emission from the vicinity of the event horizon contains significant information about the physical parameters of a black hole.

Motivated by the evidence of quasi-periodic enhanced emission within Sgr A* accretion flow (see Genzel et al. 2003[1]), the simplest model for a flare is that of a transient orbiting bright spot that would dominate the entire emission. A confirmed periodic signal from a flare could be very helpful with constraining the flaring nature and serving as evidence for strong general relativistic effects around Sgr A*.

Emitting flaring material within the accretion disk around a black hole is orbiting at high velocities, close to the speed of light, in strong gravitational potential. The emission from such flares is distorted by Doppler shifts, length contraction, time dilation, gravitational redshift and light bending. For a flare within the disk of the black hole, strong relativistic effects affect every aspect of the radiation coming from the source, including its spectrum, light-curve and image. A distant observer located at infinity will detect a time dependent light curve to which all parts of the orbiting blob contribute. These flux contributions will depend on the local emissivity but also on the relativistic effects and on the Doppler shift acting at any given position. The integration method used in the code takes into account the energy shift of the photons along the geodesics, the arrival timelags, and the lensing effect.

The Kerr metric solution implies the presence of a radius within which the orbit of a moving particle in the equatorial plane becomes unstable. This radius is known as the Innermost Stable Circular Orbit (ISCO), referred to as the marginally stable orbit (Bardeen, Press & Teukolsky 1972[2]). Beyond the ISCO, the particles plunge into the black hole on nearly geodesic orbits of constant energy and angular momentum. The ISCO represents the closest orbit to the black hole, where the orbit of a particle is stable. Below this orbit, only unstable (unbound) orbits may occur. The basic assumption is that the accretion disc is in unstable orbital motion and disc extends down to the ISCO. Most of the accretion energy is released very close to the black hole at only a few gravitational radii enabling the effects of general relativity to be probed by observation. Accreting gas can efficiently lose energy and angular momentum only outside of the ISCO. The radiative behaviour of the inner disk becomes strongly dependent on the geometry of spacetime near this region. It is very difficult to recover unambiguous information about the main parameters, such as black hole spin, disk inclination angle or inner disk radius. This is because the smooth, featureless tails in the light curve do not retain enough information to break the degeneracy between the various parameters, describing the accretion disk or its emissivity. The inner radius of the accretion disc cannot be smaller than the ISCO. This of course does not mean that there is no matter at radii lower than the ISCO but matter must spiral in (see Krolik & Hawley, 2002 [3] for alternative definitions of the "edge" of the disc).

The motion around a black hole follows bound orbits that are never closed. There are no perfectly circular orbits in a realistic scenario. Any eccentric orbit in either Schwarzschild geometry or in the equatorial plane of a Kerr black hole will be planar but never closed. Lense-Thirring precession [7] will cause for any circular orbit that lies outside the equatorial plane to not be planar and take the shape of an orbital ascending node. A plasma blob that falls into the event horizon on a trajectory outside the equatorial plane would cause the geometrically thin accretion disk to become a thick disk, symmetrically spread above and beneath the equatorial plane. The trajectory will eventually be confined onto the equatorial plane and become circular. With good approximation, we generate most of our relativistic light curves using unstable orbits in the equatorial plane. We also propose a plunging scenario of a blob in non-keplerian orbital motion, falling into the event horizon of a Kerr black hole.

We focus on recent time variability observations of Sgr A* and aim to analyze the modulation observed in Sgr A* spectrum. As the material in the accretion disk is very hot, line emission is here excluded and hence continuum emission from synchrotron radiation is only considered for modelling. The dominant radiation from a thin accretion disk is in the form of continuum emission, that provides very limited information about the environment close to the black hole.

Most models for the Sgr A* variable emission have mainly focussed on the link between the strong flares in NIR and X-ray, assuming that the IR emission is caused by synchrotron emission from a transient population of near-Gev electrons in a ∼10–100G magnetic field. The X-ray emission would be produced either by synchrotron emission from higher energy accelerated electrons or by the inverse Compton scattering of either the sub-mm emission produced in the surroundings or even directly by the synchrotron-self-Compton transient population. The transient population of accelerated electrons can also be produced via reconnection, acceleration in weak shocks or heating by plasma waves led by instabilities in the accretion flow or dissipation of magnetic turbulence. As the emission regions were proven to be compact, these models can help with understanding of the evolution of the accelerated electron populations in response to heating and radiative mechanisms (Bittner et al. 2007).

The calculations are based on a modified KY code (Dovciak, Karas, Yaqoob, 2004 [4]) for a single spot orbiting close to the corresponding last stable orbit. We adapted an existing Fortran 77 numerical code to work for broad band continuum instead of line profiles and to allow intrinsic variation of the source.

The light curves are plotted in proper time and therefore can model realistic time dependent profiles of the continuum radiation expected to be emitted by an SMBH. The code produces light curves of from a synchrotron-emitting blob of plasma. For simplicity reasons, the mapping between the rest frame emission and that seen by a distant observer is called Transfer function. Each relativistic effect has been computed for light rays emitted from the equatorial plane of the Kerr black hole and received by an observer at infinity and is stored in a separate transfer function. The transfer function includes all relativistic effects into a single function that describes the overall influence of the gravitational field on light rays emerging from the disc.

The photons (null geodesics) that connect the emission from the disc with the observer can only be found by computing the full general relativistic light travel paths which connect the disc to the observer. These null geodesics are given by solutions of the geodesic equations (Carter 1968[5]; Misner, Thorne & Wheeler 1973[6]; Chandrasekhar 1983[7]), which can be obtained numerically, but can also be given in terms of analytic functions (Rauch & Blandford 1994 [8]; Agol 1997[9]; Cadez, Fanton & Calvani 1998), which enable them to be solved quickly and with arbitrary accuracy. The first calculation of the spectrum of matter accreting onto a Kerr black hole was done by Cunningham (1975). His results, however, do not have the form suitable for data analysis, as they are tabulated only for a few values of the relevant parameters. So far, the most efficient procedure for the data analysis comes from the concept of the photon transfer function, which is constructed by calculating the trajectories of a large number of photons (e.g., Laor 1991[10]).

We use the modified KY code (publicly available) as a ray-tracer to generate time-dependent profiles of continuum emission from the black hole disk. The code requires the use of the standard NASA X-ray Spectral Fitting XSPEC Package (Arnaud, 1996). The code has been previously used by Zamaninasab et. Al (2008) to simulate the relativistic effects on Sgr A* Near-infrared light curves in order to model an evolving spot in Keplerian orbit around a Schwarzschild black hole.

The integration of photon trajectories is performed twice, both forward and backward in time, in order to compute relativistic path trajectories of the photons. All relativistic effects are pre-calculated, using forward integration subroutines (Dovciak et. al. 2004), all present in FITS files. We modify only backward integration subroutines that describe the local emission from a spot around the black hole and use the existing pre-calculated relativistic effects. All dynamical ray-tracing computations are done using photons emitted from the initial point within the accretion disk and reach the observer at infinity. The Kerr Black Hole Ray Tracer maps emitting points in the equatorial plane of a Kerr black hole to points on the observers screen. Spectral flux of continuum emission is computed by numerical integration over the solid angle subtended by the screen. All relativistic effects such as gravitational redshift, beaming and lensing are accounted for. The final formulation takes full account of all special and general relativistic effects in the photon transport and the relative motion of the emitting source. In our models the continuum emission profile originate from a thin accretion disk, the motion of the emitter in the disk being determined by the gravity and spin of the black hole as well as the space-time structure near the black hole. Time-resolved spectra of Sgr A* emissions from within the plunging region were previously discussed more succinctly in earlier papers (Howard, 2009 [11], Howard, 2010 [12], Howard, 2011 [13]).

We consider the case where disk material is within the marginally stable circular orbit at ISCO, it is assumed to be in free fall onto the black hole, with energy and angular momentum corresponding to the marginally stable orbit (Cunningham 1975 [14]; also Reynolds and Begelman 1997). All material, in free fall, is taken to lie in a slab whose thickness is small everywhere compared to the distance to the black hole (lying in thin disk). The information contained in the continuum emission profile is rich but the parameter space for the models is also fairly large.

The model parameters and light curve characteristics depend on disk radii, emissivity parameters of the source, observed inclination angle of the disk, orbital period, spin parameter of the black hole. If the black hole is rotating, we assume that the disk lies in the equatorial plane of the black hole and that the disk is corotating. This last assumption is justified by the Bardeen-Petterson (1975) effect which causes a viscous disk to stabilize in the equatorial plane of the rotating hole. With these specifications for the relativistic accretion disk system we proceed to calculate quantities relevant to our study. We compute the photon flux from a local proper area in the disc reaching the detector within a given solid angle and being measured in the rest frame co-moving with the disc. The flux contributions are integrated over a mesh grid on the disc surface. The observed radiation flux from a spot in an accretion disc is obtained by integrating the intrinsic emission over the entire spot, weighted by the transfer function determining the impact of relativistic energy change (Doppler and gravitational) as well as the lensing effect for a distant observer directed along the inclination angle.

A number of elements contribute to the observed variability of the emitting source:

* Doppler shift between the emitted and detected radiation that increase the intensity of the rays originating at the approaching side of the source, and vice versa for the receding side. The effect tends to enlarge the light curve and to enhance the observed flux at its high energy tail. The effect is expected to depend on the inclination angle of the observer. Doppler boosting effect becomes significant for high inclination angles.

* Gravitational redshift that affects mostly the photons coming from the inner parts of the source. This effect decreases the amplitude of the main peak in the light curve.

* Light-travel time that determines which photons are received at the same instant of time. Contours of constant time delay are deformed near the black hole. This effect influences the flux under suitable orientation angles.

* Gravitational lensing manifested in the distortion of the disk relative to an image in flat space. The effect enhances the flux from the far side of the source (the part of the source which is at the upper conjunction with the black hole). This tends to increase the observed flux in the light curve.

We consider a steady flat-spectrum continuum emission from a single orbiting blob of white light continuum emission spectrum. The variation in flux from a blob arises from at least two fundamental processes. The first is the frequency shifting of photons from the Doppler effect as the blob orbits the black hole. The second is the flux amplification from gravitational lensing of photons by the black hole. The presence of these two effects, along with the flight time of photons to the observer, produces the light curve. The profile of the light curve is mainly due to the motion of the hot spot in a curved spacetime around the black hole.

Motivated by the evidence of the observed periodicity in the NIR emission from Sgr A*, the simplest model for a flare is that of a transient over-dense bright compact region or a hot spot, orbiting inside an accretion flow and dominating the quiescent emission in the disk. The observed periodicities that suggest the presence of temporary clumps of matter in the accretion flow are not theoretically unexpected and may be common in the region near the ISCO, as strong inhomogeneities develop in the innermost regions of general relativistic magnetohydrodynamic simulations. Such a hot spot can be created by a flare, internal instabilities, stellar impacts or irregularities in the disk. Such a spot may also appear as the product of magnetic reconnection events or turbulent shocks within the accretion flow. The observed light curves contain significant information about the spacetime around the black hole. The hot spot model consists of a locally axially symmetric over-density of non-thermal electron bright region in circular and confined Keplerian orbital motion around the black hole. The hot spot model for continuum emission is characterized by the black hole mass, spin parameter, disk inclination angle, spot size, radial distance from the black hole, spectral index and frequency of the emitted light. The model produces a perfectly periodic light curve as a single hot spot orbits the black hole indefinitely. We assume a stationary local emission and a dependence on the axial coordinate together with the prescribed dependencies on radius, spot size, spectral index and frequency.

The model produces a QPO signal associated with the Keplerian orbit of a hot spot of plasma at the last stable orbit. Orbital periods of 20 min are of significant importance for Sgr A* flare emission study, as they correspond to a quasi-periodic (QPO) signal, interpreted as the orbit of a spot at the last stable orbit of a Kerr black hole. It is also possible that such a hot spot may appear somewhere farther away from the ISCO, producing QPOs on longer timescales. While there is evidence for periodic modulations of 20 min, a major difficulty induced by the hot spot model is the lack of significant peaks in the periodograms, while performing robust statistical estimators, therefore no conclusive results in NIR regime (Do et al. 2008, Meyer et al. 2008). Additionally, some MHD simulations estimate that a hot spot could last less than an orbital time scale before it disappears.

## 2. Orbital periodicity in Sgr A* light curves

Sgr A* exhibits strong flares in the NIR and X-ray, implying that the innermost parts of the accreting region are strongly variable. Various observations show a periodic variability varying between around 20 min and around 40 min (Genzel et al. 2003; Dodds-Eden et al. 2010). These pseudo-periodicities are interpreted as the period of an orbiting hot spot.

The measurement of orbital periodicity has the potential to place a lower limit on the black hole spin parameter (Genzel et al. 2003; Broderick & Loeb 2005; Meyer et al. 2006 [15]). There is evidence for a 17.3 minute periodic modulation in NIR and X-ray bands (Genzel et al.2003, Eckart et al. 2006 [16], Meyer et al. 2006, Belanger et al. 2006 [17], Aschenbach et al. 2004).

The 17-22 mins periodicities were observed clearly only in a few cases, raising doubts about their orbital origin, but Eckart et al. (2006) found that the QPOs are present in NIR polarized light, while still not visible in the total observed flux. In this way, it is possible that the QPOs, while invisible in the total flux, could indeed be present in all NIR flares.

As the orbital period at the last stable orbit of a Schwarzschild black hole is much longer than 20 min, this periodicity is the closest evidence for Kerr metric around Sgr A*.

A fruitful characteristic of the model is a weak dependence on the spin parameter. Most models that aim to produce QPO signals include either Keplerian orbits of hot spots at ISCO (Meyer et al., 2006; Trippe et al., 2007 [18]) or models with rotational instabilities in the accretion flow.

The orbit is stable only for radii larger than the ISCO and consequently it is dependent on the spin parameter. A lower limit of the orbital timescale is here imposed by the spin of the black hole.

For example, Sgr A* dynamical timescale lies between 27.3 min for a Schwarzschild black hole with $a$=0 and 3.7 min for an extreme black hole with $a$=1, assuming prograde orbits, $a$>0. However, a flare could actually form at radial locations larger than the ISCO.

Assuming that the detected periodic signal origin is orbital, for the co-rotating ISCO, Genzel (Genzel et.al., 2003) derives a lower limit of the spin parameter of $a$=0.52, whereas based on disk oscillation model, Aschenbach (Aschenbach, 2007) found $a$=0.996 and Kato found $a$=0.4 (Kato et.al. 2009).

Assuming the shortest period claimed to date, 13.2 min, the Sgr A* spin parameter becomes $a$>0.70.

For the flare from 2002, 3 October, Belanger et al. derived the orbital period of 22.2 min. Belanger also found the presence of a 22.2 min period in the X-ray flaring event from 2004 August 31 using simultaneous XMM-Newton and HST data (see Yusef-Zadeh et al. 2006 [19], Belanger et al. 2005 and Belanger et al. 2006).

A bright flare was simultaneously observed in NIR and X-ray on 2007, April 4 (Dodd-Eden et al. 2009, Porquet et al. 2008). The NIR power spectrum shows quasi-periodic variations of about 23 min.

Assuming the orbital period to be close to the NIR observed periodicity, of about 22 minutes, it is possible that only one of the peaks of the light curve to contribute to the orbital period of the flare and the other peak to be caused by the intrinsic variability of the flare.

From the observed X-ray flare, assuming that only one of the peaks, in this case, the second peak, is responsible for the orbital periodicity and corresponds to the lensing peak, we notice a number of important features: the presence of another peak that could correspond to the intrinsic flare and a third smaller peak on the right side of the flare, which most likely corresponds to the Doppler blueshift effect on the approaching side of the orbit.

## 3. Plunging blob model

Both NIR and X-ray observations provide evidence for a blob orbiting very close to the ISCO, on a non-Keplerian unstable orbit, falling in towards the ISCO on the last portion of the accretion disk.

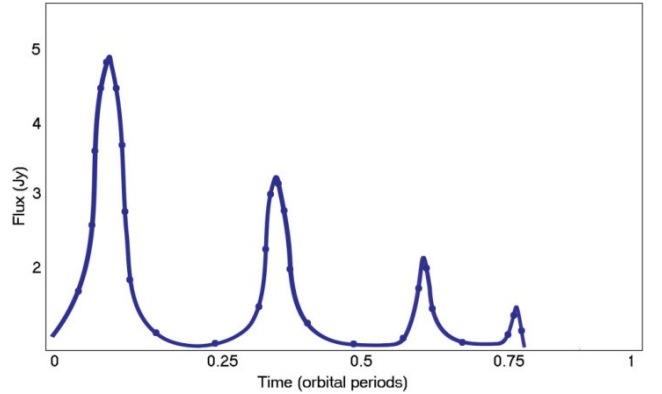

**Figure 1**: Relativistic light curve from a initially located at $r_{sp}$=0.93$GM/c^2$ falling into the event horizon of an extremely rotating black hole, at a viewing angle $\theta_o$=75°. The radius of the spot is $R_{spot}$=0.5$GM/c^2$. The orbital period decreases monotonically with time.

We use a simple in-falling flare in a plunging orbit within the ISCO. The light curve for an in-falling flare that models the 2005 X-ray light curve, can be seen in Figure 1.

The quasi-periodicity refers to the average period of a detected signal with periodic features, while there is still a possible evolution from longer to shorter periods, leading to successively broader peaks.

We model the time evolution of such an emitting compact source in plunging orbit around the black hole, with successive smaller orbital periods while the flux distribution decreases, as the blob approaches the event horizon.

However, we don't eliminate the possibility to have plunging orbits for orbital radii larger than ISCO, due to a possible evolution detected by Belanger in X-ray from longer to shorter orbital periods.

The evolution of the orbital period as a function of time was considered, as it is expected from the in-spiraling motion on a plunging orbit close to the ISCO, as seen in Figure 2.

We here show the relativistic computation of the motion of blob in an unstable (unbound) orbit around the black hole, within the marginally stable orbit.

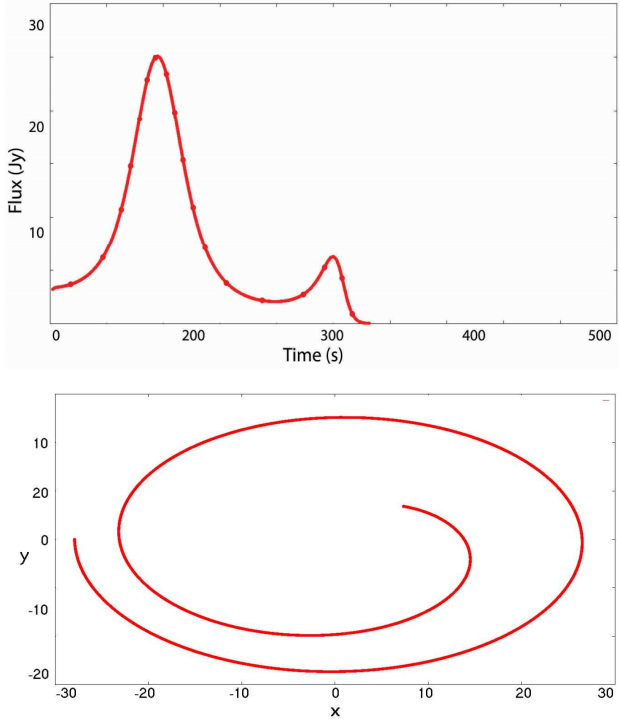

**Figure 2**: Relativistic light curve from a blob initially located at $r_{sp}$=0.85$GM/c^2$ falling into the event horizon an extremely rotating black hole, at $\theta_o$=60°. The radius of the spot is $R_{spot}$=0.2$GM/c^2$. (Top) Trajectory in the equatorial plane of the same flare. (Bottom)

The period of 17–23 minutes corresponds to orbital radii of 0.73–0.94 $r_{ISCO}$ below the ISCO ($r_{ISCO}$=3$R_s$=6$GM/c^2$) for a Schwarzschild black hole. The X-ray flare is modelled using a continuously decreasing orbital period of the orbiting material. The magnetic reconnection event could be interpreted as a sudden decrease in the magnetic field in an inner region of the disk, occurring every 20 min, on the natural dynamical timescale. Magnetic reconnection events are expected to occur near the ISCO, accompanied by sudden heating of the inner regions of the accretion flow due to magnetic dissipation.

Figure 4 shows the Belanger et. al (2008) [20] light curve for the observed August 31, 2004 flare. We show a comparison between Falanga et. al (2008) [21] best fit model, at 72° inclination angle (solid black line) with our relativistic computation of a blob in plunging orbital motion around a Kerr black hole (at the same inclination angle). Our model shows a possible evolution of the period (between 17–23 minutes) as a function of time, as expected from the spiraling motion of matter within the marginally stable orbit.

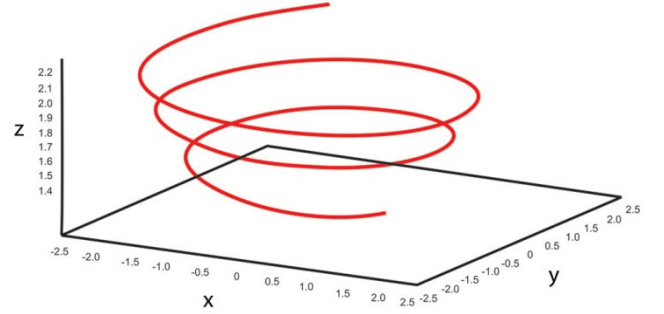

**Figure 3**: Plunging trajectory blob initially located at $r_{sp}$=0.93$GM/c^2$ on an unstable non-Keplerian orbit, falling into the event horizon of Kerr black hole with $a/M$=0.9987, at a viewing angle $\theta_o$=75°. The radius of the spot is $R_{spot}$=0.5$GM/c^2$.

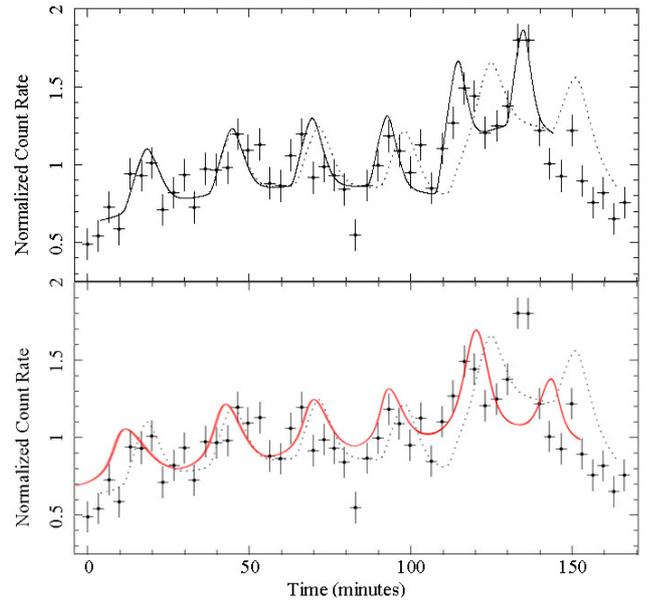

**Figure 4:** Light curve of the August 31, 2004 flare in the 2–10 keV energy band (Belanger et al. 2008), normalized with the observed mean count rate of 0.231 cts s$^{-1}$ for the flare duration. The best fit model from Falanga et. al (2008) is shown as a black solid line and our relativistic model is shown as a red solid line (inclination angle 72°). The dotted curve represents a constant Keplerian period at the last stable orbit, i.e., $r_{ISCO}$ at 72° inclination angle.

The flare emission may be produced by a hot, magnetized Keplerian flow of clumping matter (Yuan et al. 2004; Kato et al. 2009), in in-spiralling orbital motion very close to the ISCO and the stress edge of the accretion disk (Krolik and Hawley, 2002). The flare observed emission may be directly associated here with a monotonically decreasing orbital period. The decrease of the period would suggest the temporal evolution of an compact source that moves inwards across the ISCO of the black hole. The orbit is no longer Keplerian due to the presence of a radial velocity component. The X-ray flare lifetime is shorter than the NIR flare counterpart. When simultaneously detected, the X-ray flare could be most likely related to the flare infrared counterpart. The similar morphology of the sub-structure in the light curves suggests the same population of highly energetic electrons that trigger the flare emission. As the X-ray and near-IR emission could be coupled, we may expect very strong similarities in the light curves.

The trajectory of a particle in Kerr metric in Boyer-Lindquist coordinates approaching the event horizon at $x=x_h$ follows a plunging trajectory, quasi-periodically "winding up" with the angular velocity $\Omega_\varphi \rightarrow \Omega_h$, coming close to the angular velocity of the black hole horizon $\Omega_h$, where

$$\Omega_h = \frac{d\varphi}{dt}\bigg|_{x \rightarrow x_h} = \frac{a}{2(1+\sqrt{1-a^2})}. \quad (1)$$

Figure 3 shows the plunging trajectory of a flare with the angular velocity $\Omega_\varphi \rightarrow \Omega_h$, located between the event horizon and the ISCO, quasi-periodically infalling into the black hole and approaching the event horizon at the southern hemisphere. The dependence of the angular velocity on the orbital radius, for a plunging particle or photon from the ISCO into the event horizon is shown in Figure 5. We show both Schwarzschild and Kerr black hole cases. The angular velocity of black hole event horizon $\Omega_h$ may be imprinted to the QPO variable signal from the black hole.

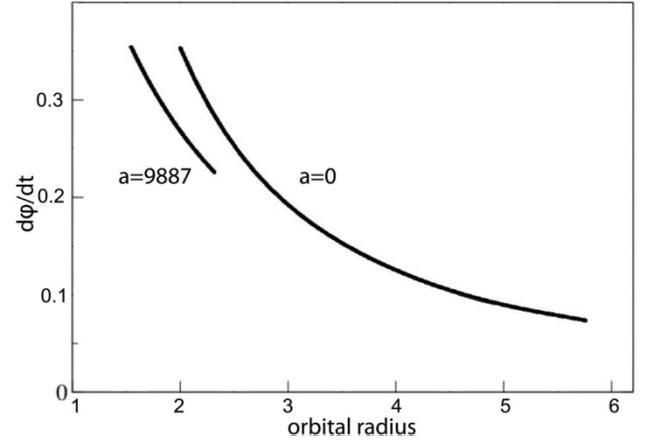

**Figure 5**: The angular velocity evolution during the plunge of the blob from ISCO towards the event horizon, for an extremely rotating black hole and for a Schwarzschild black hole.

However the NIR and X-ray Sgr A* observations are limited to 1-2 hours flare event durations. Due to this short duration, the QPO periodicity is still open to argument. However, the QPOs are only seen at the flaring time and most likely occur near the ISCO (Genzel et al. (2003); Aschenbach et al. (2004); Eckart et al. (2006); Belanger et al. (2006)). The blob approaches the event horizon and it is viewed by the observer as a relativistically beamed and boosted forward emission that is repeated quasi-periodically with a frequency very close to $\nu_h = \Omega_h/2\pi$. An oscillation with the angular frequency $\Omega_h$ could be a proper observational signature for the spin of the black hole and may be related to the QPOs of non-equatorial bound orbits in the accretion flow.

## 4. Conclusion

We developed a relativistic time-dependent plunging model takes into account all special and general relativistic effects. We sought for relativistic signatures in the resulting light curves. We computed the profile of the resulting light curve and the blending between Doppler shifts, lensing, relativistic beaming, and gravitational redshifting.

We found that the variation in flux would arise from two major fundamental processes: Doppler shift and gravi-tational lensing. The resulting light curve corresponds to a local event close to the ISCO.

Future simultaneous measurements at both optically thin and thick frequencies should be able to test the model. A further understanding of Sgr A* emission will rely on simultaneous multi-wavelength campaigns. The flare emission is most likely originating from a compact source from the inner regions of the accretion disk and could probe the dynamics and properties of the gravitational field close to an event horizon.

We have studied and modelled the possibility of gravitational lensing and Doppler effects in light curves from compact orbiting blobs. The VLTI instrument GRAVITY will be able to image these plasma blobs orbiting near the ISCO. VLBI interferometry or second-generation Very Large Telescope Interferometer instrument GRAVITY is designed to detect highly relativistic motion of matter close to the event horizon. The astrometric displacement of the primary image, the presence of a secondary image that shifts the centroid of Sgr A, and the gravitational lensing effect may be detected. GRAVITY may also detect gravitational lensing effects in the NIR bands. The gravitational lensing effects around Sgr A* might be detected for the first time due to GRAVITY astrometric capacity.

While the model offers a realistic picture of Sgr A* variability at several wavelengths and predicts several features in the flare emission, there are still a number of issues to be addressed. Several approximations had to be assumed in order to keep the calculations simple. In Kerr metric, a good approximation is neglecting the gravitational field of the accreting plasma. If the disc is thick and dense, its gravity, especially farther away from the black hole, shouldn't be neglected. A more advanced version of the model that assumes a self-gravitating disc should be considered. We also assumed that the blob is compact and rigid, neglecting any shearing effects. In a realistic scenario, the blob could vary in shape due to differential rotation within the accretion disk. In a very strong magnetic field, the blob can be considered a rigid object that retains its original shape during the flare emission event. The blob intrinsic shape is disregarded as the source is ray traced from infinity to the disk, using the geometric optics approximation. We have also ignored external influences on the light rays when they arrive at the observer. The blob follows a stable planar circular orbit near the ISCO in the equatorial plane of the disk.

A supermassive black hole is unlikely to be accreting in the same orbital plane as the black hole spin. The Lense-Thirring precession should lead to a tilted and warped disk. Eccentric orbits should eventually converge and produce non-axisymmetric standing shocks. The gravitational torque could cause the disc to precess almost as a solid body (Fragile et al. 2007). By working in 2 dimensions only, the simulation excludes the effects of misalignment between the angular momentum of the gas and the black hole. A geometrically thick accretion flow does not align with the black hole. The disc misalignment affects the dynamics and properties of the accretion flow. A tilted accretion flow could alter the observable features of Sgr A* emission and considerably affect the sub-flare structure of the light curve. In our modeling the flare is caused by a local event close to the ISCO. In such a dynamical picture of the accretion flow, it is possible that the flaring to be produced close but not exactly at the ISCO due to instabilities and turbulent processes in the disk. For a thick disk solution, self-eclipsing of the disk might play some role in the observed emission. As the blobs could be created through magneto-rotational instabilities, shown to be present in Keplerian rotating accretion disks (Hawley Balbus 1991), a shear-flow instability may generate a turbulent flow in the disk. Excess angular momentum transport from such shock could truncate the accretion flow outside the ISCO (Fragile 2009; Dexter Fragile 2011). The flare could spread out via a shock wave and produce a temporary torus around the black hole. In this case, the spin of the black hole shouldn't be derived only based on flares located at ISCO. The light curve sub-structure could be also affected by the presence of an orbiting asymmetry in this torus causing flux modulations on orbital scales.

We examined the contributions of relativistic effects associated with the plunging motion to the variability of Sgr A*. Using ray-tracing techniques, we modelled all relativistic signatures in light curves from a continuum emitting blob in plunging orbital motion around a Kerr black hole.

We tested the theoretical model in a simplistic and phenomenological way, in order to resist to a qualitative comparison with simultaneous light curves in future observation campaigns.